\documentclass{ijuc}
\usepackage{array}
\usepackage{multirow}
\usepackage{amsmath}
\usepackage{amssymb}
\usepackage[pdftex]{graphicx}

\usepackage{booktabs}
\usepackage{braket}

\usepackage{bbm}
\usepackage{microtype}

\usepackage{hyperref}
\usepackage{doi}

\begin{document}

\title{Boolean logic gate design principles in unconventional computers:
an NMR case study}

\author{
Matthias Bechmann\inst{1}\email{matthias.bechmann@york.ac.uk}
\and
Angelika Sebald\inst{1}\email{angelika.sebald@york.ac.uk}
\and
Susan Stepney\inst{2}\email{susan@cs.york.ac.uk}
}

\institute{
Department of Chemistry, University of York, YO10 5DD, York, UK
\and
Department of Computer Science, University of York, YO10 5DD, York, UK
}

\def\received{Received 5 September 2011}

\maketitle

\begin{abstract}
We present a general method for analysing novel computational substrates
to determine which of their parameters can be manipulated to exhibit
the complete set of 2-input boolean logical operations. We demonstrate
this approach with an NMR-based case study, showing which NMR parameters
can be used to perform boolean logic.
\end{abstract}

\keywords{NMR, boolean logic, canalising functions, universal gates, spin}

\section{Introduction}

There is much work on \emph{in materio} computing: exploiting unconventional
material substrates and their dynamical properties to perform computation,
and examining their computational properties and abilities. One of
the more advanced is optical computing \cite{Woods2008,Woods2009}.
Other, more exotic, substrates include nuclear spins in NMR experiments
\cite{Jones2001,Jones2011,Rosello-Merino2010,BechmannICES2010}, liquid
crystals \cite{Harding,Miller,Harding06}, conductive media \cite{Mills_2008,Mills_al_2006},
reaction-diffusion chemical systems \cite{Kuhnert_1989,Motoike2005107,Sielewiesiuk_2001,Toth_1995},
DNA \cite{Adleman11111994,Amos}, and even slime moulds \cite{Adamatzky_2007,Tero22012010}.

The aim is to allow the material to do what comes naturally, under
control of physical laws, and to exploit this natural dynamics as
an efficient form of computation \cite{SS-PhysicaD-08,SS-BCS-VCS,SS-UC11}.
The question arises: given some novel substrate, how to analyse its
properties to determine whether it is a suitable medium for computation,
and, if so, how it can be manipulated to perform computation in a
manner best suited to that substrate.

Here we tackle a part of that problem: analysing a novel substrate
to determine how it can be used to perform boolean logical operations.
We present the design principles through a case study of using bulk
nuclear spins, in the context of NMR (nuclear magnetic resonance)
experiments.

\section{Theory and Background}

\subsection{Boolean functions and logic gates }

The NAND and NOR logic gates are both \emph{universal}, in that either
can be used to construct all the other boolean 2-input logic gates.
However, to do so, several such gates may be required. In an implementation,
it is often more important to minimise the circuitry (the number of
actual gates) than to minimise the number of types of gates.

Logic gates can be classified in terms of their symmetry properties
\cite{Slepian1953,Golomb1959}. In terms of minimising circuitry,
gates can be classified in terms of equivalence classes on permutations
(rewiring) of inputs, and on negation of inputs and/or outputs (i.e.\
adding inverters, important in cases where they are significantly
\emph{cheaper} to implement than binary gates \cite{Correia2001}).
There are four such equivalence classes for the 16 boolean 2-input
logic gates (Tables~\ref{tab:bool-equiv} and \ref{tab:bool-egs}).

\begin{table}
\centering
\begin{tabular}{cl}
\toprule
class & members\tabularnewline
\midrule
0 & T, F\tabularnewline
1 & $A$, NOT $A$, $B$, NOT $B$\tabularnewline
2 & $A$ AND $B$; $A$ NAND $B$; NOT $A$ AND $B$, $A$ AND NOT $B$,
\ldots{}\tabularnewline
3 & $A$ XOR $B$, NOT ($A$ XOR $B$)\tabularnewline
\bottomrule
\end{tabular}
\caption{\label{tab:bool-equiv}The four equivalence classes of the 16 boolean
2-input logic gates. These correspond to equivalence classes under
permutations of inputs and negation of the inputs and/or outputs \cite{Correia2001},
and to the different kinds of canalising functions \cite{Drossel2009}. }
\end{table}

\begin{table}
\centering
\begin{tabular}{cc|cccc}
\toprule
$A$ & $B$ & T & $B$ & $A$ NAND $B$ & $A$ XOR $B$\tabularnewline
\midrule
0 & 0 & 1 & 0 & 1 & 0\tabularnewline
0 & 1 & 1 & 1 & 1 & 1\tabularnewline
1 & 0 & 1 & 0 & 1 & 1\tabularnewline
1 & 1 & 1 & 1 & 0 & 0\tabularnewline
\bottomrule
\end{tabular}
\caption{\label{tab:bool-egs}Truth tables of examples of the four classes
of boolean logic gates, for each possible input value of $A$ and
$B$. Class 0 (example T) is independent of either input value.
Class 1 (example $B$) is independent of one of the input values (here,
$A$) for all possible inputs. Class 2 (example $A$ NAND $B$) is
independent of one of the input values when the other has a particular
value (here, it it independent of the value of $B$ when $A$ has
the value $0$). Class 3 (example $A$ XOR $B$) depends on both input
values for all inputs.}
\end{table}

The same classification is found by considering canalising functions
\cite{Drossel2009}. These are functions where the output is independent
of one (or more) of the inputs, for one (or more) input values. Class
0 consists of the constant functions: the output is completely independent
of the inputs. Class 1 consists of the strongly canalising functions:
the output is independent of one of the inputs (for example, the function
$B$ is independent of the value of input $A$). Class 2 consists
of the weakly canalising functions: the output is independent of one
of the inputs when the other input has a specific value (for example,
the function $A$ NAND $B$ is independent of the value of input $B$
when the value of input $A$ is false). Class 3 consists of the non-canalising
functions: the output is determined by both inputs.

\begin{figure}
	\begin{centering}
		\includegraphics[width=1\columnwidth]{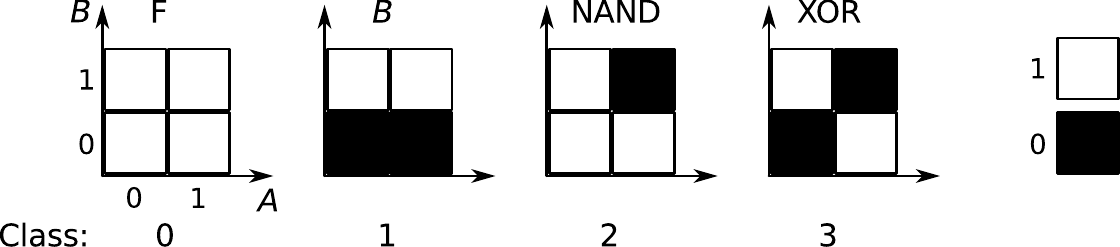}
		\par\end{centering}
		
		\caption{\label{fig:Logic-gates-patterns}Logic-gates patterns and their different
			symmetries as visible from the two-dimensional representations of
			the canalising functions gate classes 0, 1, 2 and 3.}
		\end{figure}

We use these symmetries/canalising properties of the gates (Figure~\ref{fig:Logic-gates-patterns})
to look for analogous properties exhibited by the substrate that indicates a natural implementation route.

\subsection{Nuclear spin dynamics}

The most common application of NMR spectroscopy is that as a tool
for structure elucidation of condensed matter in general, and of molecules
in the liquid state in particular. The quantised nature of nuclear
spin can also be assigned to the notion of a qubit in quantum
computations \cite{Schumacher1995}. Nuclear spins have also been
used as a substrate to implement classical computation paradigms such
as binary or continuous logic gates and circuits \cite{Rosello-Merino2010,BechmannICES2010}. 

A major reason for these computational applications to choose nuclear
magnetic spins as the implementation platform is the rigour by which
nuclear spin dynamics are described by quantum mechanics. In addition,
the accuracy by which the macroscopically measured signal in an NMR
experiment can be related to the underlying microscopic spin dynamics,
using a density matrix approach, is nearly unrivalled by any other
spectroscopic technique. In order to develop the formal relationship
between the properties of logic gates and the quantum mechanical expressions
describing the spin dynamics in a NMR experiment, we give a short
summary of the necessary formalisms. For more details see textbooks on NMR \cite{Levitt2008}.
Here we consider only uncoupled spins $S=1/2$ in the liquid state.

\begin{figure}
\begin{centering}
\includegraphics[width=0.4\textwidth]{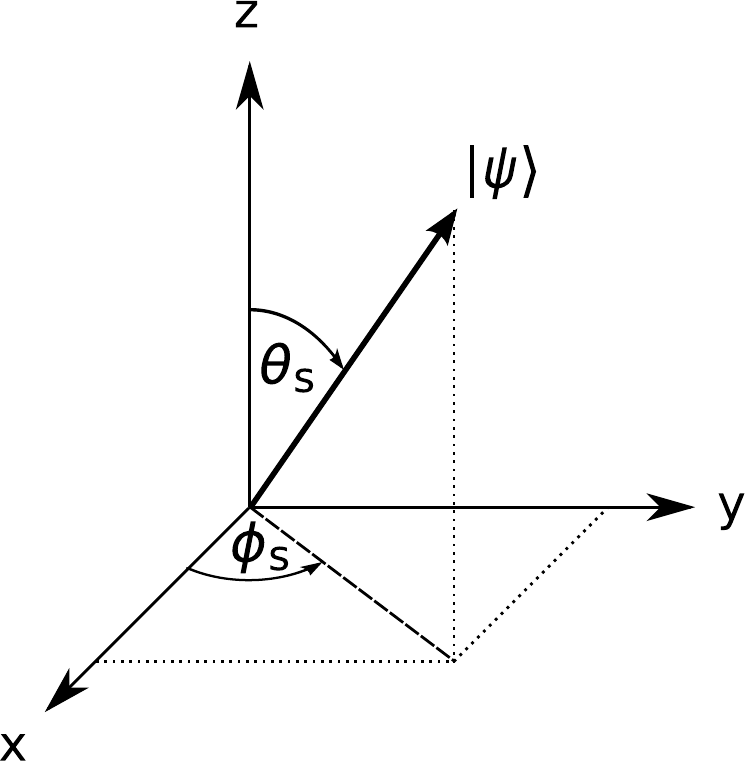}
\par\end{centering}

\caption{\label{fig:Spin-state-vector-Cartesian}Spin state vector orientation
of $\Ket{\psi}$ in its Cartesian representation}
\end{figure}

The time evolution of an initial spin state vector $\Ket{\psi\left(t_{a}\right)}$
is described as 
\begin{eqnarray}
\Ket{\psi\left(t\right)} & = & \mathcal{U}\left(t,t_{a}\right)\Ket{\psi\left(t_{a}\right)}\label{eq:spin-ket-evolution}
\end{eqnarray}
where $\mathcal{U}\left(t,t_{a}\right)$ is the time propagator describing
the spin dynamics at every given point in time. The general orientation
of the spin vector $\Ket{\psi}$ is \cite{Levitt2008} 
\begin{equation}
\Ket{\psi}=\left(\begin{array}{c}
\cos\frac{\theta_{\mathrm{s}}}{2}\mathrm{e}^{-\mathrm{i}\frac{1}{2}\phi_{\mathrm{s}}}\\
\sin\frac{\theta_{\mathrm{s}}}{2}\mathrm{e}^{+\mathrm{i}\frac{1}{2}\phi_{\mathrm{s}}}
\end{array}\right)\label{eq:Spin-Vector-Orientation}
\end{equation}
where $\theta_{\mathrm{s}}$ and $\phi_{\mathrm{s}}$ are the polar and
azimuth angles. Its three-dimensional representation is given by 
\begin{equation}
\left(\begin{array}{c}
\Braket{\psi|I_{x}|\psi}\\
\Braket{\psi|I_{y}|\psi}\\
\Braket{\psi|I_{z}|\psi}
\end{array}\right)=\frac{1}{2}\left(\begin{array}{c}
\sin\theta_{\mathrm{s}}\cos\phi_{\mathrm{s}}\\
\sin\theta_{\mathrm{s}}\sin\phi_{\mathrm{s}}\\
\cos\theta_{\mathrm{s}}
\end{array}\right)\label{eq:Spin-Vector-Orientation-3D}
\end{equation}
describing a general Cartesian vector orientation (see Figure~\ref{fig:Spin-state-vector-Cartesian}),
where $I_{x}$, $I_{y}$ and $I_{z}$ are basis spin operators \cite{Sakurai1994}.

The time propagation of the macroscopic NMR signal is described by
the density matrix $\rho(t)=\overline{\Ket{\psi\left(t\right)}\Bra{\psi\left(t\right)}}$
as 
\begin{eqnarray}
\rho(t) & = & \mathcal{U}\left(t,t_{a}\right)\rho(t_{a})\mathcal{U}^{\dagger}\left(t,t_{a}\right)
\end{eqnarray}
where the bar signals the ensemble average. The thermal equilibrium
density matrix is \cite{Levitt2008} 
\begin{equation}
\rho_{z}^{\mathrm{eq}}=\frac{1}{2}\mathbbm{1}+\frac{1}{2}\lambda_{\mathrm{B}}I_{z}\label{eq:thermal-eq-rho}
\end{equation}
where $\lambda_{\mathrm{B}}=\hbar\gamma B_{0}/(k_{\mathrm{B}}T)$
defines a Boltzmann factor, scaling the separation of spin energy
levels, that can be interpreted as a spin polarisation along the $z$-axis.

\begin{figure}
\begin{centering}
\includegraphics[width=1\columnwidth]{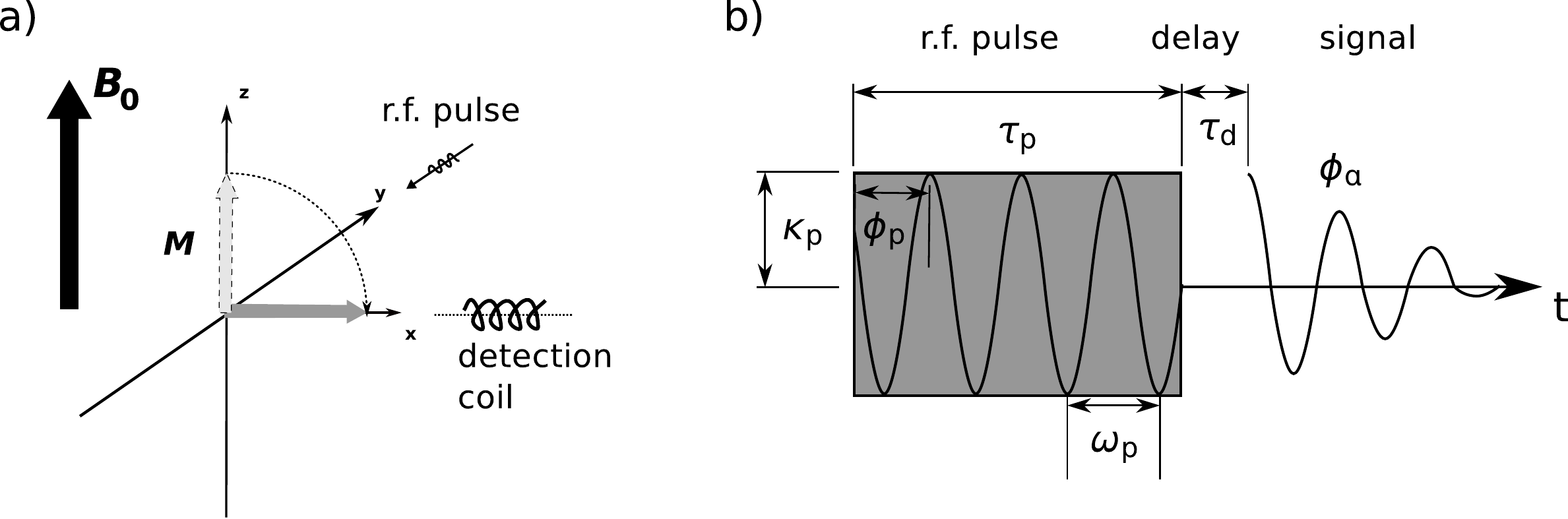}
\par\end{centering}

\caption{\label{fig:NMR-pulse-experiment}NMR single-pulse experiment: (a)
rotation of the magnetisation vector $\boldsymbol{M}$ by a r.f.\
pulse $R_{\phi_{\mathrm{p}}}\left(\beta\right)$ with $\phi_{\mathrm{p}}=\frac{\pi}{2}$
and $\beta=\kappa_{\mathrm{p}}\tau_{\mathrm{p}}=\frac{\pi}{2}$; (b)
the relevant parameters describing the single-pulse NMR experiment
(pulse amplitude $\kappa_{\mathrm{p}}$, pulse duration $\tau_{\mathrm{p}}$,
pulse phase $\phi_{\mathrm{p}}$, pulse frequency $\omega_{\mathrm{p}}$,
free evolution delay $\tau{}_{\mathrm{d}}$ and acquisition phase
$\phi_{\mathrm{a}}$).}
\end{figure}

The NMR system Hamiltonian and generator of the time evolution propagators
for isolated, uncoupled spins is 
\begin{equation}
\mathcal{H}=\mathcal{H}_{\mathrm{CS}}+\mathcal{H}_{\mathrm{\mathrm{rf}}}
\end{equation}
where $\mathcal{H}_{\mathrm{CS}}$ and $\mathcal{H}_{\mathrm{rf}}$
are the chemical shielding and radio frequency (r.f.)\ Hamiltonian,
respectively. The explicit representation of the r.f.\ Hamiltonian
is particularly simple in the rotating reference frame (RRF) as
\begin{eqnarray}
\mathcal{H}_{\mathrm{rf}} & = & \omega_{\mathrm{p}}I_{z}+\kappa_{\mathrm{p}}\left(I_{x}\cos\phi_{\mathrm{p}}+I_{y}\sin\phi_{\mathrm{p}}\right)\label{eq:rfHamiltonian-RRF}
\end{eqnarray}
where $\omega_{\mathrm{p}}$ is a frequency offset (relative to
the frequency of RRF), and $\kappa_{\mathrm{p}}$ and $\phi_{\mathrm{p}}$
are the amplitude and phase of the pulse. Figure~\ref{fig:NMR-pulse-experiment}(a)
summarises the behaviour of the magnetisation vector $\boldsymbol{M}$
(Eqs.~\eqref{eq:spin-ket-evolution}--\eqref{eq:Overall-Magnetisation})
under the influence of a r.f.\ pulse; Figure~\ref{fig:NMR-pulse-experiment}(b)
depicts the relevant parameters. 
 Pulses applied to spins resonating at the rotation frequency of the
RRF are described by the system Hamiltonian 
\begin{eqnarray}
\mathcal{H}=\mathcal{H}_{\mathrm{rf}} & = & \kappa_{\mathrm{p}}\left(I_{x}\cos\phi_{\mathrm{p}}+I_{y}\sin\phi_{\mathrm{p}}\right)\label{eq:rfHamiltonian-RRF-OnRes}
\end{eqnarray}
and the overall system Hamiltonian for uncoupled isolated spins $S=1/2$
is 
\begin{eqnarray}
\mathcal{H} & = & \mathcal{H}_{\mathrm{CS}}+\mathcal{H}_{\mathrm{\mathrm{rf}}}:\mbox{during pulse}\nonumber \\
\mathcal{H} & = & \mathcal{H}_{\mathrm{CS}}\,:\mbox{during free evolution}
\end{eqnarray}
including a chemical shielding offset $\mathcal{H}_{\mathrm{CS}}$
from the RRF frequency.

The general form of the time evolution propagator in NMR is 
\begin{eqnarray}
\mathcal{U}\left(t_{b},t_{a}\right) & = & \mathcal{T}\exp\left\{-\mathrm{i}\int_{t_{a}}^{t_{b}}\mathcal{H}\left(t\right)\mathrm{d}t\right\}
\end{eqnarray}
where $\mathcal{T}$ is the Dyson time-ordering operator \cite{Sakurai1994}.

Dependent on the symmetry of the Hamiltonian, different (simpler) expressions
for the propagator can be formulated
\begin{eqnarray}
\mathcal{H}\left(t\right)=\mathcal{H} & \rightarrow & \mathcal{U}\left(t_{b},t_{a}\right)=\exp\left\{ -\mathrm{i}\mathcal{H}\left(t_{b}-t_{a}\right)\right\}\label{eq:Htimeindep}\\
 & & \mathrm{time\ independent}\nonumber\\
\left[\mathcal{H}\left(t'\right),\mathcal{H}\left(t''\right)\right]=0 & \rightarrow & \mathcal{U}\left(t_{b},t_{a}\right)=\exp\{ -\mathrm{i}{\textstyle \int}_{t_{a}}^{t_{b}}\mathrm{d}t'\mathcal{H}\left(t'\right)\} \label{eq:Hselfcom}\\
 & & \mathrm{self-commuting}\nonumber\\
\mathcal{H}\left(t_{a}+\tau\right)=-\mathcal{H}\left(t_{b}-\tau\right) & \rightarrow & \mathcal{U}\left(t_{b},t_{a}\right)=1\:\mbox{; anti-symmetric}\label{eq:Uantisym}\\
 &  & \mathcal{U}\left(t_{b},t_{a}\right)=\pm1\:\mbox{; cyclic}\label{eq:Ucyclic}
\end{eqnarray}

In the case of a hard pulse, a so-called $\delta$ pulse is usually
a valid description of the time propagator 
\begin{eqnarray}
\mathcal{U}\left(t_{b},t_{a}\right) & = & \exp\left\{ -\mathrm{i}\mathcal{H}_{\mathrm{rf}}\tau_{\mathrm{p}}\right\} ;\,\tau_{\mathrm{p}}=t_{b}-t_{a}\nonumber \\
 & = & \exp\left\{ -\mathrm{i}\kappa_{\mathrm{p}}\tau_{\mathrm{p}}\left(I_{x}\cos\phi_{\mathrm{p}}+I_{y}\sin\phi_{\mathrm{p}}\right)\right\} \nonumber \\
 & = & R_{\phi_{\mathrm{p}}}\left(\beta\right);\,\beta=\kappa_{\mathrm{p}}\tau_{\mathrm{p}}
\end{eqnarray}
which is equivalent to the rotation operation $R_{\phi_{\mathrm{p}}}\left(\beta\right)$.
It generates a rotation by an angle $\beta$ about an axis in the
$xy$-plane with azimuth angle $\phi_{\mathrm{p}}$. For a sequence
of $\delta$ pulses the propagator  can be factorised as
\begin{eqnarray}
\mathcal{U}\left(t_{c},t_{a}\right) & = & \mathcal{U}_{2}\left(t_{c},t_{b}\right)\mathcal{U}_{1}\left(t_{b},t_{a}\right);\,\tau_{\mathrm{p1}}=t_{b}-t_{a}\,\mbox{and }\tau{}_{\mathrm{p2}}=t_{c}-t_{b}\nonumber \\
 & = & R_{\phi{}_{\mathrm{p}2}}\left(\beta_{2}\right)R_{\phi_{\mathrm{p}1}}\left(\beta_{1}\right);\,\beta_{1}=\kappa_{\mathrm{p}1}\tau_{\mathrm{p}1};\,\beta_{2}=\kappa{}_{\mathrm{p}2}\tau{}_{\mathrm{p2}}
\end{eqnarray}
The explicit form of these propagators is equivalent to rotation operators
in two dimensions and is \cite{Levitt2008}

\begin{eqnarray}
R_{x}\left(\beta\right) & = & \mathrm{e}^{-\mathrm{i}\beta I_{x}}=\left(\begin{array}{cc}
\cos\frac{\beta}{2} & -\mathrm{i}\sin\frac{\beta}{2}\\
-\mathrm{i}\sin\frac{\beta}{2} & \cos\frac{\beta}{2}
\end{array}\right)\\
R_{y}\left(\beta\right) & = & \mathrm{e}^{-\mathrm{i}\beta I_{y}}=\left(\begin{array}{cc}
\cos\frac{\beta}{2} & -\sin\frac{\beta}{2}\\
\sin\frac{\beta}{2} & \cos\frac{\beta}{2}
\end{array}\right)\\
R_{z}\left(\beta\right) & = & \mathrm{e}^{-\mathrm{i}\beta I_{z}}=\left(\begin{array}{cc}
\mathrm{e}^{-\mathrm{i}\frac{\beta}{2}} & 0\\
0 & \mathrm{e}^{+\mathrm{i}\frac{\beta}{2}}
\end{array}\right)\\
R_{\phi}\left(\beta\right) & = & \mathrm{e}^{-\mathrm{i}\beta\left(I_{x}\cos\phi+I_{y}\sin\phi\right)}=\left(\begin{array}{cc}
\cos\frac{\beta}{2} & -\mathrm{i}\sin\frac{\beta}{2}\mathrm{e}^{-\mathrm{i}\phi}\\
-\mathrm{i}\sin\frac{\beta}{2}\mathrm{e}^{+\mathrm{i}\phi} & \cos\frac{\beta}{2}
\end{array}\right)\nonumber \\
 & = & R_{z}\left(\phi\right)R_{x}\left(\beta\right)R_{z}\left(-\phi\right)\label{eq:RFPulseGeneral}
\end{eqnarray}
After a pulse $R_{\phi_{\mathrm{p}}}\left(\beta\right)$ with phase
$\phi_{\mathrm{p}}=\frac{\pi}{2}$ and flip angle $\beta=\frac{\pi}{2}$
the thermal equilibrium density matrix $\rho_{z}^{\mathrm{eq}}$ is
transformed according to 
\begin{eqnarray}
R_{\frac{\pi}{2}}\left(\frac{\pi}{2}\right)\rho_{z}^{\mathrm{eq}}R_{\frac{\pi}{2}}^{\dagger}\left(\frac{\pi}{2}\right) & =\frac{1}{2}\mathbbm{1}+\frac{\lambda_{\mathrm{B}}}{2} & \left(\begin{array}{cc}
0 & 1\\
1 & 0
\end{array}\right)\nonumber \\
 & = & \frac{1}{2}\mathbbm{1}+\frac{1}{2}\lambda_{\mathrm{B}}I_{x}\nonumber \\
 & = & \rho_{x}\label{eq:rho-x}
\end{eqnarray}
yielding spin polarisation along the $+x$ direction.

A r.f.\ pulse $R_{\phi_{\mathrm{p}}}\left(\beta\right)$ applied
to an initial state $\rho^{\left(i\right)}$ generates a spin density
matrix that can be expressed in terms of its final polar coordinates
$\rho\left(\phi_{\mathrm{s}},\theta_{s}\right)=\overline{\Ket{\psi}\Bra{\psi}}$
(Eq.~\eqref{eq:Spin-Vector-Orientation}) or as a function of the
pulse parameters rotating an initial state $\rho^{\left(i\right)}\left(\phi_{\mathrm{p}},\beta\right)$.
For example when starting from $\rho^{\left(z\right)}=\rho_{z}^{\mathrm{eq}}$
the final state is
\begin{eqnarray}
\rho^{\left(z\right)}\left(\phi_{\mathrm{p}},\beta\right) & = & \frac{1}{2}\mathbbm{1}+\frac{\lambda_{\mathrm{B}}}{2}\left(\begin{array}{cc}
\cos\beta & -\mathrm{i}\sin\beta\mathrm{e}^{-\mathrm{i}\phi_{\mathrm{p}}}\\
-\mathrm{i}\sin\beta\mathrm{e}^{+\mathrm{i}\phi_{\mathrm{p}}} & -\cos\beta
\end{array}\right)\label{eq:rho-pulse-param}
\end{eqnarray}
and the relation $\rho\left(\phi_{\mathrm{s}},\theta_{s}\right)=\rho^{\left(z\right)}\left(\phi_{\mathrm{p}},\beta\right)$
holds for $\left(\phi_{\mathrm{s}},\theta_{s}\right)=\left(\phi_{\mathrm{p}}-\frac{\pi}{2},\beta\right)$.

Equivalently to the calculation of the three-dimensional spin-vector
orientation in Eq.~\eqref{eq:Spin-Vector-Orientation-3D} from the
two-dimensional spin state $\Ket{\psi}$ (Eq.~\eqref{eq:Spin-Vector-Orientation}),
the three-dimensional magnetisation vector $\boldsymbol{M}$ can then
be calculated from any given two-dimensional density matrix $\rho$
as
\begin{eqnarray}
\boldsymbol{M} & = & \left(\begin{array}{c}
\overline{\Braket{I_{x}}}\\
\overline{\Braket{I_{y}}}\\
\overline{\Braket{I_{z}}}
\end{array}\right)=\left(\begin{array}{c}
\mbox{Tr}\left\{ \rho I_{x}\right\} \\
\mbox{Tr}\left\{ \rho I_{y}\right\} \\
\mbox{Tr}\left\{ \rho I_{z}\right\} 
\end{array}\right)\label{eq:Overall-Magnetisation}
\end{eqnarray}
In the case of single isolated spins this also implies the equivalence
in interpretation of the spin vector orientation in Figure~\ref{fig:Spin-state-vector-Cartesian}
and the magnetisation vector $\boldsymbol{M}$. 

In general a NMR pulse sequence can consist of any number and parametrisation
of r.f.\ pulses and free evolution delays. This provides for numerous
options for the implementation of e.g.\ logic gates. The most useful
parametrisation in the following is the one describing the NMR signal
as a function of the pulse sequence, $\rho^{\left(i\right)}\left(\phi_{\mathrm{p}},\beta\right)$.

\section{Logic Gates from NMR parameters}

\subsection{Classification of NMR parameters }

The most common parameters describing a NMR experiment, open for use
as control parameters to implement logic operations, are $\left\{ \kappa_{\mathrm{p}},\phi_{\mathrm{p}},\tau_{\mathrm{p}},\omega_{\mathrm{p}},\tau_{\mathrm{d}},\phi_{\mathrm{a}}\right\} $:
pulse amplitude, pulse phase, pulse duration, pulse frequency, free
evolution delay, and receiver phase, respectively (see Figure~\ref{fig:NMR-pulse-experiment}). 

A heuristic classification rule has been given elsewhere \cite{Rosello-Merino2010},
separating logic gates based on the relation between the experimental
NMR parameters chosen:
\begin{enumerate}
\item if the effect of the first parameter cannot be compensated for by
setting the second parameter (canalising input value), then $\mbox{AND},>,<\mbox{NOR},\mbox{OR},\leq,\geq,\mbox{NAND}$
gates can be constructed (asymmetric gates);
\item if the effect of the first parameter can be compensated by a setting
of the second parameter, then XOR and XNOR gates can be constructed
(symmetric gates).
\end{enumerate}
This classification of logic gates seamlessly integrates in a generalised
description by using the concept of canalising functions~ \cite{Drossel2009}
and equivalence classes \cite{Correia2001}. Fixing one of the inputs
of a logic gate ($A$ or $B$) is a \emph{canalising input value}
if changing the other (unfixed) input does not alter the gate output.
The number of all possible canalising input values for a given logic
gate for all possible inputs states is used as a criterion to assign
logic gates to one of four classes (see Table~\ref{tab:Canalised-operations-class}):
\begin{table}
\centering
\begin{tabular}{clcc}
\toprule
class & canalising input values for fixed: & $A$ & $B$\tabularnewline
\midrule
0 & T & 2 & 2\tabularnewline
1 & $B$ & 0 & 2\tabularnewline
2 & $A$ NAND $B$ & 1 & 1\tabularnewline
3 & $A$ XOR $B$ & 0 & 0\tabularnewline
\bottomrule
\end{tabular}
\caption{\label{tab:Canalised-operations-class}Number of canalising input
values for the four gate classes. Demonstrated for one representative
member of every class: T, $B$, $A$ NAND $B$, $A$ XOR $B$. The
respective truth tables for these four gates are shown in Table~\ref{tab:bool-egs}.}
\end{table}

\begin{description}
\item [{class~0}] two canalising input values for either of the two inputs
$A$ and $B$
\item [{class~1}] two canalising input values for exactly one of the inputs
$A$ or $B$
\item [{class~2}] one canalising input values for either of the two inputs
$A$ and $B$
\item [{class~3}] no canalising input values for either inputs $A$ and $B$
\end{description}

These four classes also show unique symmetry patterns in their two-dimensional
representations (see Figure~\ref{fig:Logic-gates-patterns}). Rows
and columns of equal coloured squares signal the presence of a canalising
input. These patterns can be used to map functions $f(x1,x2)$
to corresponding canalising functions and, therefore, the logic gate(s)
they can naturally implement are readily identified.

According to our previous definitions \cite{Rosello-Merino2010} a
strong parameter would generate a canalising function for at least
one of its states (class 0, 1 and 2), because altering the other parameter
causes no change in the NMR output. On the other hand, for a weak
parameter the output would always change (class 3).

In the following we demonstrate how the abstract concept of canalising
functions relates to the symmetries and commutation relations of the
NMR Hamiltonian, the time evolution propagator, and the experimental
output of a NMR experiment.

\subsection{Canalising input values in NMR context}

Here we consider a system Hamiltonian that 
consists only of the $\mathcal{H}_{\mathrm{rf}}$ term. Hence, we are in
the single spin, strong pulse, on-resonant regime, and for simplicity
we only consider the experimental parameters $\kappa_{\mathrm{p}},\phi_{\mathrm{p}},\tau_{\mathrm{p}}$
(r.f.\ pulse amplitude, phase and duration) for logic gate
generation.

The NMR r.f.\ pulse propagator $R_{\phi_{\mathrm{p}}}\left(\beta\right)=R_{z}\left(\phi_{\mathrm{p}}\right)R_{x}\left(\beta\right)R_{z}\left(-\phi_{\mathrm{p}}\right)$\\
(Eq.~\eqref{eq:RFPulseGeneral}) is the general rotation operator
about an axis in the traverse $xy$-plane of the RRF. It therefore
commutes with a particular spin state ($I_{\phi_{\mathrm{p}}}$) in
the transverse plane for a given value $\phi_{\mathrm{p}}$ (e.g.\
for $\phi_{\mathrm{p}}=0$; $\left[R_{0}(\beta),I_{x}\right]=0$ independent
of the value of $\beta$). However, $R_{\phi_{\mathrm{p}}}\left(\beta\right)$
will never commute with the thermal equilibrium state $\rho_{z}^{\mathrm{eq}}$
($\sim I_{z}$) (Eq. \eqref{eq:thermal-eq-rho}), which is perpendicular
to the $xy$-plane. For example, a $R_{0}\left(\frac{\pi}{2}\right)$
pulse applied to $z$-magnetisation ($\sim I_{z}$) will flip it to
the $-y$ direction, and therefore change the spin state and its orientation.
A subsequent pulse applied using $R_{-\frac{\pi}{2}}\left(\frac{\pi}{2}\right)$
will leave it unaltered, since the system is in the $-I_{y}$ spin
state, an eigenstate of $R_{-\frac{\pi}{2}}\left(\frac{\pi}{2}\right)$
and, therefore, a canalising input is generated in this second step.
One can postulate now that a \emph{canalising input}\textbf{ }is generated
by an r.f.\ pulse if:
\begin{enumerate}
\item $R_{\phi_{\mathrm{p}}}\left(\beta\right)$ leaves the system state
unaltered (eigenstate;$\left[R_{\phi_{\mathrm{p}}}(\beta),\rho\right]=0$),
\item $R_{\phi_{\mathrm{p}}}\left(\beta\right)=\mathbbm{1}$, the unity
operator.
\end{enumerate}
In the following this behaviour and the possibility of generating canalising
input values is analysed. First we consider single-pulse experiments
assuming different initial system-state preparations $\rho^{\left(i\right)}$.
Second we consider experiments composed of more than one pulse.

\subsubsection{Single-pulse gates}

The signal as detected during a NMR experiment is not the effect of
the full spin magnetisation vector $\boldsymbol{M}$ as given in Eq.~\eqref{eq:Overall-Magnetisation}
but only its projection into the $xy$-plane \cite{Levitt2008}. Hence,
the physically meaningful quantities that can be measured by NMR are
the magnetisation vector components $M_{x}=\overline{\Braket{I_{x}}}$
and $M_{y}=\overline{\Braket{I_{y}}}$, describing the orientational
distribution of this projection, and the magnitude $M_{xy}=\sqrt{\Braket{I_{x}}^{2}+\Braket{I_{y}}^{2}}$
describing the total amount of magnetisation present in the $xy$-plane.
The functional structure of $M_{x}$, $M_{y}$ and $M_{xy}$ is
determined by the r.f.\ pulse $R_{\phi_{\mathrm{p}}}\left(\beta\right)$ 
and the initial spin state $\rho^{\left(i\right)}$ to which the r.f. pulse is applied.
$M_{x}$, $M_{y}$ and $M_{xy}$ are therefore functions of pulse phase $\phi_{\mathrm{p}}$, pulse flip angle
$\beta=\kappa_{\mathrm{p}}\tau_{\mathrm{p}}$ and the initial direction of the 
spin magnetisation.

The spin magnetisation can assume every possible orientation, while
the r.f.\ pulse and therefore the rotation axis of the spin magnetisation
is restricted to the $xy$-plane. It is instructive to examine scenarios
where the initial spin state $\rho^{\left(i\right)}$ is either perpendicular
to the $xy$-plane and never commutes with the r.f.\ pulse operator
$R_{\phi_{\mathrm{p}}}\left(\beta\right)$, or where it is coplanar
to it and therefore can commute.

\paragraph{Starting from thermal equilibrium state }

\begin{figure}
\begin{centering}
\includegraphics[width=1\columnwidth]{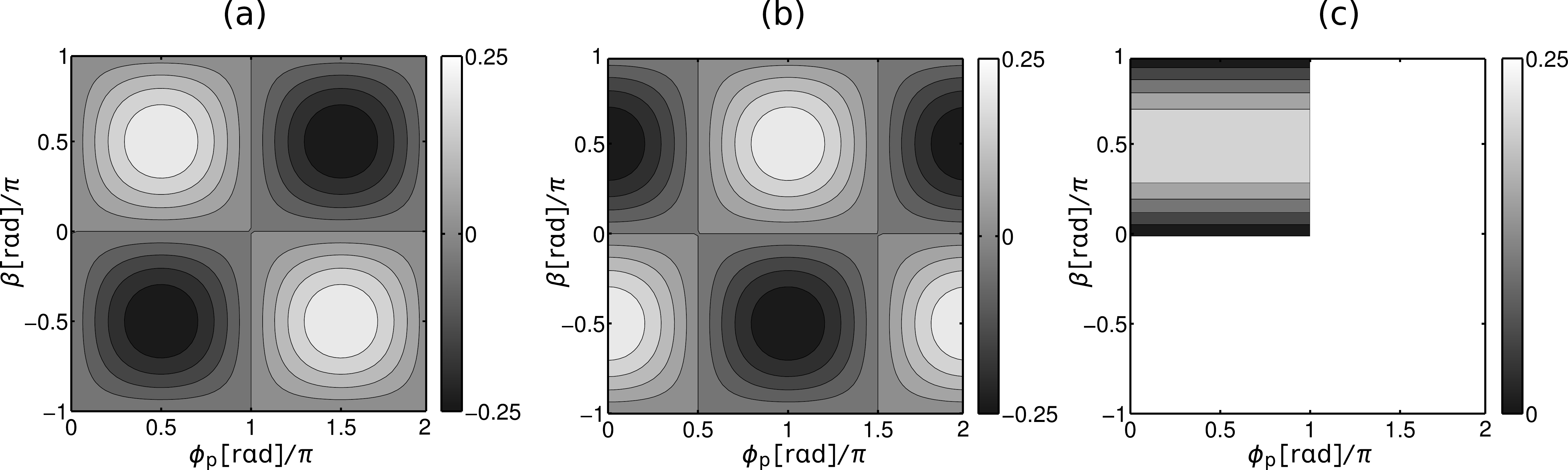}
\par\end{centering}

\caption{\label{fig:Single-pulse-measurments-Iz}Contour plots of the NMR detectable
quantities (a) $M_{x}=\overline{\Braket{I_{x}}}$, (b)$M_{y}=\overline{\Braket{I_{y}}}$
and (c) $M_{xy}=\sqrt{\overline{\Braket{I_{x}}}^{2}+\overline{\Braket{I_{y}}}^{2}}$
(Eq.~\eqref{eq:thermal-eq-single-pulse}) as a function of the pulse
parameters $\phi_{\mathrm{p}}$ and $\beta$. The initial spin polarisation
has been along the $z$-axis ($\rho_{z}^{\mathrm{eq}}$) and $\lambda_{\mathrm{B}}=1$
has been assumed. The ranges for $\phi_{\mathrm{p}}$ and $\beta$
have been chosen such that the only symmetry operation necessary to
generate a full, infinite plot are horizontal and vertical translations.}
\end{figure}

The most simple and natural initial spin state is the one the system
assumes at thermal equilibrium $\rho^{\left(i\right)}=\rho_{z}^{\mathrm{eq}}$
(Eq.~\eqref{eq:thermal-eq-rho}). Here the spin polarisation is pointing
along the $z$-axis, perpendicular to the $xy$-plane and $R_{\phi_{\mathrm{p}}}(\beta)$
never commutes with it. $\rho_{z}^{\mathrm{eq}}$ transforms under
the influence of a pulse $R_{\phi_{\mathrm{p}}}(\beta)$ as 
\begin{eqnarray}
\rho^{\left(z\right)}\left(\phi_{\mathrm{p}},\beta\right) & = & R_{\phi_{\mathrm{p}}}(\beta)\rho_{z}^{\mathrm{eq}},R_{\phi_{\mathrm{p}}}^{\dagger}(\beta)
\end{eqnarray}
After this single pulse the spin magnetisation vector $\boldsymbol{M}$
(Eq.~\eqref{eq:Overall-Magnetisation}) is a function of pulse flip
angle $\beta$ and pulse phase $\phi_{\mathrm{p}}$ 
\begin{eqnarray}
\left(\begin{array}{c}
M_{x}\\
M_{y}\\
M_{z}
\end{array}\right)  =  \left(\begin{array}{c}
\mathrm{Tr}\left\{ \rho^{\left(z\right)}(\phi_{\mathrm{p}},\beta)I_{x}\right\} \\
\mathrm{Tr}\left\{ \rho^{\left(z\right)}(\phi_{\mathrm{p}},\beta)I_{y}\right\} \\
\mathrm{Tr}\left\{ \rho^{\left(z\right)}(\phi_{\mathrm{p}},\beta)I_{z}\right\} 
\end{array}\right)=\frac{\lambda_{\mathrm{B}}}{4}\left(\begin{array}{c}
\sin\phi_{\mathrm{p}}\sin\beta\\
-\cos\phi_{\mathrm{p}}\sin\beta\\
\cos\beta
\end{array}\right)\label{eq:thermal-eq-single-pulse}
\end{eqnarray}
and the NMR measurable quantities $M_{x}\left(\phi_{\mathrm{p}},\beta\right)$,
$M_{y}\left(\phi_{\mathrm{p}},\beta\right)$ and $M_{xy}\left(\phi_{\mathrm{p}},\beta\right)=\frac{\lambda_{\mathrm{B}}}{4}\left|\sin\beta\right|$
are readily determined. These quantities are functions of the two
pulse parameters $\phi_{\mathrm{p}}$ and $\beta$ just like the binary
logic gates are functions of the two inputs $A$ and $B$ (Table~\ref{tab:bool-egs}).
An instructive way to analyse $M_{x}\left(\phi_{\mathrm{p}},\beta\right)$,
$M_{y}\left(\phi_{\mathrm{p}},\beta\right)$ and $M_{xy}\left(\phi_{\mathrm{p}},\beta\right)$,
as to which logic gates can be implemented by them, is by examining
their representation as two-dimensional contour plots. These are shown
in Figure~\ref{fig:Single-pulse-measurments-Iz}. 
 A comparison with the two-dimensional sketches of 2-input logic gates
in Figure~\ref{fig:Logic-gates-patterns} immediately reveals agreements
and disagreements in symmetries. 

In principle the functions $M_{x}\left(\phi_{\mathrm{p}},\beta\right)$,
$M_{y}\left(\phi_{\mathrm{p}},\beta\right)$ and $M_{xy}\left(\phi_{\mathrm{p}},\beta\right)$
are continuous in $\phi_{\mathrm{p}}$ and $\beta$ whereas logic
gates are boolean functions of $A$ and $B$ that can only assume
the discrete values of $\left\{ 0,1\right\} $. In order to map the
continuous functions in Figure~\ref{fig:Single-pulse-measurments-Iz}
to the four possible discrete sets ($A$,$B$) of a logic gate one
has to find four discrete pairs $\left(\phi_{\mathrm{p}},\beta\right)$
at which to evaluate the continuous functions and where the symmetry
pattern of the desired logic gate results.

For example from Figure~\ref{fig:Single-pulse-measurments-Iz}(a)
depicting $M_{x}$, one can directly identify the symmetry pattern
corresponding to a class 3 gate in Figure~\ref{fig:Logic-gates-patterns}.
In order to implement the class 3 (XOR) gate (Table~\ref{tab:bool-egs})
one needs to define four discrete value sets $\left(\phi_{\mathrm{p}}=\phi_{\mathrm{p}}^{A},\beta=\beta^{B}\right)$
(with $A$,$B$ $\in\{0,1\}$). Inspection of Figure~\ref{fig:Single-pulse-measurments-Iz}(a)
immediately suggests the positive and negative extrema as possible
candidates: Selecting the pair $\left(\phi_{\mathrm{p}}^{A=0},\beta^{B=0}\right)=\left(\frac{\pi}{2},-\frac{\pi}{2}\right)$
the detected output is $M_{x}\left(\phi_{\mathrm{p}}^{A=0}=\frac{\pi}{2},\beta^{B=0}=-\frac{\pi}{2}\right)=-0.25$.
In the same fashion, $M_{x}\left(\frac{\pi}{2},\frac{\pi}{2}\right)=0.25$,
$M_{x}\left(\frac{3\pi}{2},-\frac{\pi}{2}\right)=0.25$ and $M_{x}\left(\frac{3\pi}{2},\frac{\pi}{2}\right)=-0.25$
are calculated for the remaining XOR gate input configurations. The
values of $M_{x}\left(\phi_{\mathrm{p}},\beta\right)$, $M_{y}\left(\phi_{\mathrm{p}},\beta\right)$
and $M_{xy}\left(\phi_{\mathrm{p}},\beta\right)$ in Figure~\ref{fig:Single-pulse-measurments-Iz}
are calculated assuming $\lambda_{\mathrm{B}}=1$. Mapping these values
to the final boolean values $\left\{ 0,1\right\} $ is achieved by
scaling them by a factor of 4. 

\begin{table}
\begin{tabular}{c|cc|cc|cccc|l}
\toprule
\multirow{2}{*}{class} & \multicolumn{2}{c|}{$\phi_{\mathrm{p}}^{A}$} & \multicolumn{2}{c|}{$\beta^{B}$} & \multicolumn{4}{c|}{Output=$M_{x}\left(\phi_{p}^{A},\beta^{B}\right)$} & \multirow{2}{*}{gate}\tabularnewline
\cline{2-9} 
 & 0 & 1 & 0 & 1 & 00 & 01 & 10 & 11 & \tabularnewline
\midrule
0 & $\frac{\pi}{2}$ & $\frac{5\pi}{2}$ & $\frac{\pi}{2}$ & $\frac{5\pi}{2}$ & 0.25  & 0.25 & 0.25 & 0.25 & T\tabularnewline
1 & $\frac{\pi}{2}$ & $\frac{5\pi}{2}$ & $-\frac{\pi}{2}$ & $\frac{\pi}{2}$ & -0.25 & 0.25  & -0.25  & 0.25 & $B$\tabularnewline
2 & $\pi$ & $\frac{3\pi}{2}$ & $0$ & $\frac{\pi}{2}$ & 0 & 0 & 0 & -0.25 & $A$ NAND $B$\tabularnewline
3 & $\frac{\pi}{2}$ & $\frac{3\pi}{2}$ & $-\frac{\pi}{2}$ & $\frac{\pi}{2}$ & -0.25  & 0.25  & 0.25  & -0.25 & $A$ XOR $B$\tabularnewline
\bottomrule
\end{tabular}
\caption{\label{tab:Example-value-thermal-eq-gates}Some selected values $\phi_{p}$
and $\beta$ to generate one representative gate of every gate class
using $M_{x}\left(\phi_{\mathrm{p}},\beta\right)$ as shown in Figure~\ref{fig:Single-pulse-measurments-Iz}(a).}
\end{table}

A class 3 gate is characterised by the absence of any canalising input
(Table~\ref{tab:Canalised-operations-class}). A change in $\phi_{\mathrm{p}}^{A}$
or $\beta^{B}$ will therefore always change the value $M_{x}\left(\phi_{\mathrm{p}}^{A},\beta^{B}\right)$.
In order to change one of the parameters $\phi_{\mathrm{p}}^{A}$
or $\beta^{B}$ and achieve the same result in $M_{x}\left(\phi_{\mathrm{p}}^{A},\beta^{B}\right)$
one can take advantage of the periodicity of the trigonometric functions
in Eq.~\eqref{eq:thermal-eq-single-pulse} and in this way $M_{x}\left(\frac{\pi}{2},-\frac{\pi}{2}\right)=M_{x}\left(\frac{5\pi}{2},-\frac{\pi}{2}\right)=-0.25$
and $M_{x}\left(\frac{\pi}{2},\frac{\pi}{2}\right)=M_{x}\left(\frac{5\pi}{2},\frac{\pi}{2}\right)=0.25$
can be implemented. This gate is evaluating $\phi_{\mathrm{p}}^{A}$
at values separated by $2\pi$ and the gate is therefore independent
of a change in $\phi_{\mathrm{p}}^{A}$ no matter which of its two
permitted values $\beta^{B}$ assumes. A change in $\beta^{B}$, however,
will always change $M_{x}\left(\phi_{\mathrm{p}}^{A},\beta^{B}\right)$.
This behaviour corresponds to a class 1 gate with two canalising input
values for one parameter. In order to implement class 2 gates a comparison
of the logic gate patterns and the contour plot for $M_{x}\left(\phi_{\mathrm{p}}^{A},\beta^{B}\right)$
suggests a restriction of the parameter range to smaller intervals.
For example, one may choose the interval $\phi_{\mathrm{p}}\in\left[\pi,\frac{3\pi}{2}\right]$
and $\beta\in\left[0,\frac{\pi}{2}\right]$ to implement a NAND gate
representing a class 2 (NAND) gate (see Table~\ref{tab:Example-value-thermal-eq-gates}).
The implementation of a class 0 gate requires two canalising input
values for both $\phi_{\mathrm{p}}^{A}$ and $\beta^{B}$. As in the
class 1 ($B$) gate scenario the periodicity of $M_{x}\left(\phi_{\mathrm{p}}^{A},\beta^{B}\right)$
can be used to achieve canalising inputs. However, this time both
parameters have to undergo the $2\pi$ value changes. Alternatively
one can use pulses corresponding to rotation operators $R_{\phi_{\mathrm{p}}}\left(\beta\right)=\pm\mathbbm{1}$
for generating canalising inputs for any parameter configuration of
$\phi_{\mathrm{p}}^{A}$ or $\beta^{B}$.

Having demonstrated how to implement one member of every gate class
it is now a trivial task to generate implementations of all the members
of a given gate class. By using the symmetry operations of permuting
the input assignment of $A$ and $B$ to $\phi_{\mathrm{p}}$ and
$\beta$, inverting the input or the output, the remaining gates are
obtained directly.

This shows that all 2-input logic gates can be implemented using the
functions $M_{x}\left(\phi_{\mathrm{p}},\beta\right)$. $M_{x}\left(\phi_{\mathrm{p}},\beta\right)$
and $M_{y}\left(\phi_{\mathrm{p}},\beta\right)$ are both products
of two linear trigonometric functions, differing only by a phase shift
of $\frac{\pi}{2}$ in the factor depending on $\phi_{\mathrm{p}}$
($\sin\left(\phi_{\mathrm{p}}+\frac{\pi}{2}\right)=\cos\phi_{\mathrm{p}}$).
This phase shift can be seen in Figure~\ref{fig:Single-pulse-measurments-Iz}(a)
and (b) as a horizontal shift by $\frac{\pi}{2}$, otherwise both
plots are identical. Their overall symmetry pattern is that of a class
3 gate. Hence, everything said about $M_{x}\left(\phi_{\mathrm{p}},\beta\right)$
implementations equally holds for $M_{y}\left(\phi_{\mathrm{p}},\beta\right)$. 

However the same can not be said about the magnetisation magnitude
$M_{xy}\left(\phi_{\mathrm{p}},\beta\right)$ (Figure~\ref{fig:Single-pulse-measurments-Iz}(c)).
$M_{xy}\left(\phi_{\mathrm{p}},\beta\right)=M_{xy}\left(\beta\right)=\frac{\lambda_{\mathrm{B}}}{4}\left|\sin\beta\right|$
is only a function of the pulse flip angle $\beta$ and, therefore,
displays a contour plot with a symmetry corresponding to a class 1
gate. 

The behaviour of the functions under permutation of the variables
$\phi_{\mathrm{p}}$ and $\beta$ and/or their inversion distinguishes
$M_{x}\left(\phi_{\mathrm{p}},\beta\right)$, $M_{y}\left(\phi_{\mathrm{p}},\beta\right)$
and $M_{xy}\left(\phi_{\mathrm{p}},\beta\right)$ from each other.
$M_{x}\left(\phi_{\mathrm{p}},\beta\right)$ is invariant to permutation
of the variables while $M_{xy}\left(\phi_{\mathrm{p}},\beta\right)$
is not. In contrast, $M_{x}\left(\phi_{\mathrm{p}},\beta\right)$
changes under inversion of its variables $\phi_{\mathrm{p}}\rightarrow-\phi_{\mathrm{p}}$
or $\beta\rightarrow-\beta$, while $M_{xy}\left(\phi_{\mathrm{p}},\beta\right)$
is invariant to inversion of $\phi_{\mathrm{p}}\rightarrow-\phi_{\mathrm{p}}$.
$M_{y}\left(\phi_{\mathrm{p}},\beta\right)$ shows the same behaviour
as $M_{xy}\left(\phi_{\mathrm{p}},\beta\right)$ under these transformations.
However, a simple shift in $\phi_{\mathrm{p}}=\frac{\pi}{2}$ gives
$M_{x}\left(\phi_{\mathrm{p}},\beta\right)=M_{y}\left(\phi_{\mathrm{p}}+\frac{\pi}{2},\beta\right)$,
an operation which is not possible for $M_{xy}\left(\phi_{\mathrm{p}},\beta\right)$
since it is independent of $\phi_{\mathrm{p}}$. The differences in
behaviour arise because for $M_{x}\left(\phi_{\mathrm{p}},\beta\right)$
and $M_{y}\left(\phi_{\mathrm{p}},\beta\right)$ both variables $\phi_{\mathrm{p}}$
and $\beta$ are arguments of products of trigonometric functions,
which generates function values in the range of $\left[-1,1\right]$,
whereas $M_{xy}\left(\phi_{\mathrm{p}},\beta\right)$ has trigonometric
function values in $\left[-1,1\right]$ only for the variable $\beta$.

In principle, the task of implementing all 2-input logic gates by NMR
spectroscopy is already accomplished by using only the simplest
of all NMR experiments: starting from thermal equilibrium and only
using a single r.f.\ pulse. This scenario can therefore serve as
a Universal Logic Module (ULM) \cite{Yau1970}. There are, however,
good reasons why one needs to explore other starting conditions, and
more complicated NMR pulse sequences. NMR is uniquely suitable as
a single testbed for the implementation of classical as well as quantum
computations. In common formulations of quantum algorithms the initial
state of the computation is not the thermal equilibrium state but
a superposition state \cite{Jones2001}. For valid comparisons between
classical and quantum algorithm NMR implementations the same initial
(superposition) state should be used. Further, it may be desirable
to construct more extended circuitry than just a single logic gate
\cite{Hardy2001}. Then gate implementations that can deal with initial
states other than $\rho_{z}^{\mathrm{eq}}$ are attractive as an efficient
means of taking advantage of the output of a logic gate without having
to restore the initial state $\rho_{z}^{\mathrm{eq}}$ before the
next logic gate in a circuit can be executed.

\paragraph{Starting from superposition state}

\begin{figure}
\begin{centering}
\includegraphics[width=1\columnwidth]{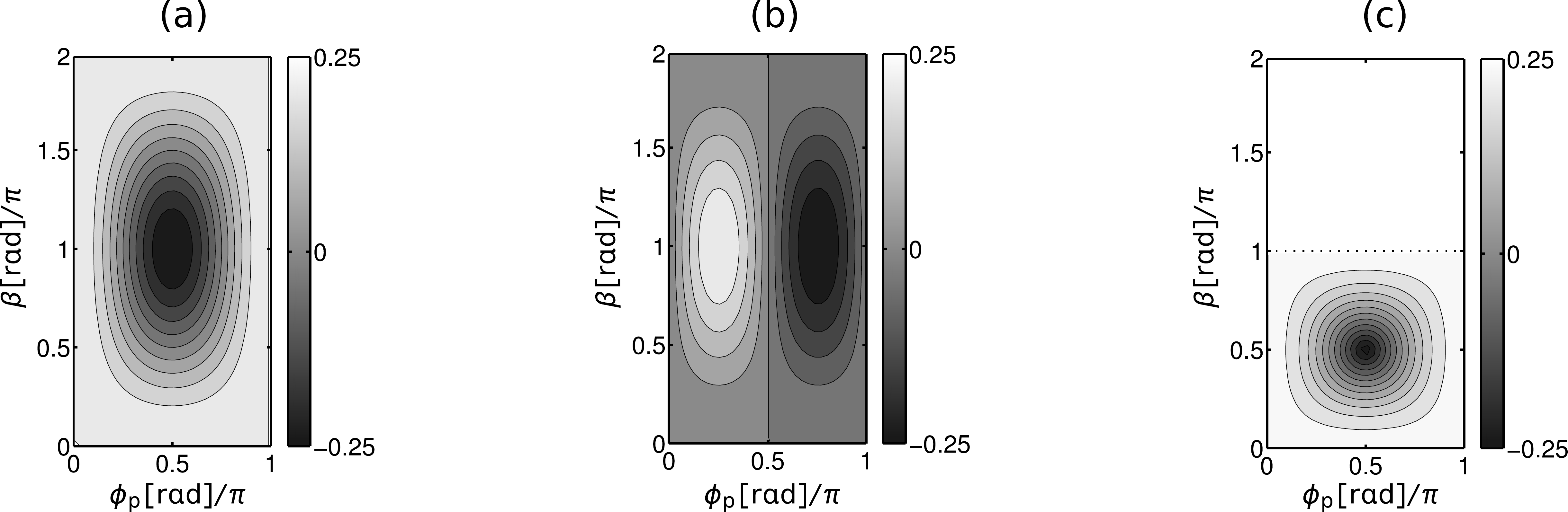}
\par\end{centering}

\caption{\label{fig:Single-pulse-measurments-Ix}Contour plots of the NMR detectable
quantities (a) $M_{x}=\overline{\Braket{I_{x}}}$, (b)$M_{y}=\overline{\Braket{I_{y}}}$
and (c) $M_{xy}=\sqrt{\overline{\Braket{I_{x}}}^{2}+\overline{\Braket{I_{y}}}^{2}}$
(Eq.~\eqref{eq:superposition-state-single pulse}) as a function
of the pulse parameters $\phi_{\mathrm{p}}$ and $\beta$. The initial
spin polarisation has been along the $x$-axis ($\rho_{x}$) and $\lambda_{\mathrm{B}}=1$
has been assumed. The ranges for $\phi_{\mathrm{p}}$ and $\beta$
have been chosen such that the only symmetry operation necessary to
generate a full, infinite plot are horizontal and vertical translations.}
\end{figure}

A r.f.\ pulse $R_{\frac{\pi}{2}}(\frac{\pi}{2})$ applied to $\rho_{z}^{\mathrm{eq}}$
generates the superposition state $\rho_{x}=\frac{1}{2}\mathbbm{1}+\frac{1}{2}\mathbb{\lambda_{\mathrm{B}}}I_{x}$, 
which corresponds to spin polarisation pointing along the $+x$-axis.
Here we take $\rho_{x}$ as the initial superposition state.
The commutator between $R_{\phi_{\mathrm{p}}}(\beta)$ and $\rho_{x}$
is 
\begin{equation}
\left[R_{\phi_{\mathrm{p}}}(\beta),\rho_{x}\right]=-\lambda_{\mathrm{B}}I_{z}\sin\phi_{\mathrm{p}}\sin\frac{\beta}{2}\label{eq:singel-pulse-commutator-superposition}
\end{equation}
 and is zero for values $\phi_{\mathrm{p}}=n\pi$. These
angles correspond to rotations around the positive and negative $x$-axis
(the commutator is also zero for $\beta=2n\pi$, but these
angles represent trivial $2\pi$ rotations). $\rho_{x}$ transforms
under the influence of a r.f.\ pulse $R_{\phi_{\mathrm{p}}}(\beta)$
as 
\begin{eqnarray}
\rho^{\left(x\right)}\left(\phi_{\mathrm{p}},\beta\right) & = & R_{\phi_{\mathrm{p}}}(\beta)\rho_{x}R_{\phi_{\mathrm{p}}}^{\dagger}(\beta)\label{eq:single-pulse-tranformation-superposition}
\end{eqnarray}
 The r.f.\ pulse generates a magnetisation vector $\boldsymbol{M}$
(Eq.~\eqref{eq:Overall-Magnetisation}) as a function of pulse flip
angle $\beta$ and pulse phase $\phi_{\mathrm{p}}$ according to 
\begin{eqnarray}
\left(\begin{array}{c}
M_{x}\\
M_{y}\\
M_{z}
\end{array}\right)  =  \left(\begin{array}{c}
\mathrm{Tr}\left\{ \rho^{\left(x\right)}(\phi_{\mathrm{p}},\beta)I_{x}\right\} \\
\mathrm{Tr}\left\{ \rho^{\left(x\right)}(\phi_{\mathrm{p}},\beta)I_{y}\right\} \\
\mathrm{Tr}\left\{ \rho^{\left(x\right)}(\phi_{\mathrm{p}},\beta)I_{z}\right\} 
\end{array}\right)=\frac{\lambda_{\mathrm{B}}}{4}\left(\begin{array}{c}
1-2\sin^{2}\phi_{\mathrm{p}}\sin^{2}\frac{\beta}{2}\\
\sin 2\phi_{\mathrm{p}}\sin^{2}\frac{\beta}{2}\\
-\sin\phi_{\mathrm{p}}\sin\beta
\end{array}\right)\label{eq:superposition-state-single pulse}
\end{eqnarray}
All the quantities $M_{x}\left(\phi_{\mathrm{p}},\beta\right)$,
$M_{y}\left(\phi_{\mathrm{p}},\beta\right)$ and\\ $M_{xy}\left(\phi_{\mathrm{p}},\beta\right)=\frac{\lambda_{\mathrm{B}}}{4}\sqrt{1-\sin^{2}\phi_{\mathrm{p}}\sin^{2}\beta}$
are functions of $\phi_{\mathrm{p}}$ and $\beta$. Contour plots
of these three functions are shown in Figure~\ref{fig:Single-pulse-measurments-Ix}.
 Again, agreements and disagreements in symmetries are found by comparison
with Figure~\ref{fig:Logic-gates-patterns}. Inspection of Figures~\ref{fig:Single-pulse-measurments-Iz}
and \ref{fig:Single-pulse-measurments-Ix} immediately reveals differences
in the transformation behaviour of $\rho_{x}$ and $\rho_{z}^{\mathrm{eq}}$. 

As shown earlier it is possible to implement any logic gate from a
NMR measurable quantity that has the symmetry of a class 3 gate in
Figure~\ref{fig:Logic-gates-patterns} (e.g. $M_{x}\left(\phi_{\mathrm{p}},\beta\right)$
in Figure~\ref{fig:Single-pulse-measurments-Iz}(a)). However, none
of the three contour plots in Figure~\ref{fig:Single-pulse-measurments-Ix}
displays this symmetry and no class 3 gates can be implemented. The
symmetries present belong to those of class 0, 1 and 2 gates and can
be found in any of the three contour plots. Implementing class 1 gates
requires taking advantage of the periodicity of the functions $M_{x}\left(\phi_{\mathrm{p}},\beta\right)$,
$M_{y}\left(\phi_{\mathrm{p}},\beta\right)$ and $M_{xy}\left(\phi_{\mathrm{p}},\beta\right)$
in order to generate canalising inputs. Class 2 gates use a reduced
parameter range for $\phi_{\mathrm{p}}$ and $\beta$. 

Permutation of $\phi_{\mathrm{p}}$ and $\beta$ changes $M_{x}\left(\phi_{\mathrm{p}},\beta\right)$,
$M_{y}\left(\phi_{\mathrm{p}},\beta\right)$, while $M_{xy}\left(\phi_{\mathrm{p}},\beta\right)$
stays invariant under this operation. $M_{x}\left(\phi_{\mathrm{p}},\beta\right)$
and $M_{xy}\left(\phi_{\mathrm{p}},\beta\right)$ are invariant under
input inversion $\phi_{\mathrm{p}}\rightarrow-\phi_{\mathrm{p}}$
or $\beta\rightarrow-\beta$ while $M_{y}\left(\phi_{\mathrm{p}},\beta\right)$
is only invariant under $\beta\rightarrow-\beta$. All functions are
now periodic with $\pi$ in $\phi_{\mathrm{p}}$ because of their
dependence on $\sin^{2}\phi_{\mathrm{p}}$ or $\sin 2\phi_{\mathrm{p}}$.
$M_{x}\left(\phi_{\mathrm{p}},\beta\right)$ and $M_{y}\left(\phi_{\mathrm{p}},\beta\right)$
have a $2\pi$ periodicity interval in $\beta$ and $\pi$ periodicity
for $M_{xy}\left(\phi_{\mathrm{p}},\beta\right)$. All the functions
$M_{x}\left(\phi_{\mathrm{p}},\beta\right)$, $M_{y}\left(\phi_{\mathrm{p}},\beta\right)$
and $M_{xy}\left(\phi_{\mathrm{p}},\beta\right)$ are the products
of two trigonometric functions of which at least one is quadratic.
The squaring reduces the trigonometric function values to the interval
$\left[0,1\right]$ and therefore, it is not possible to have a symmetry
behaviour required for class 3 gates.

The rotation operator $R_{\phi_{\mathrm{p}}}(\beta)$ \emph{not commuting}
with the initial density matrix $\rho_{z}^{\mathrm{eq}}$ leads to
the functions $M_{x}\left(\phi_{\mathrm{p}},\beta\right)$, $M_{y}\left(\phi_{\mathrm{p}},\beta\right)$
and $M_{xy}\left(\phi_{\mathrm{p}},\beta\right)$ described by products
of linear trigonometric functions and the ability to implement \emph{all}
logic gates. The rotation operator $R_{\phi_{\mathrm{p}}}(\beta)$
\emph{commuting} with the initial density matrix $\rho_{x}$, however,
leads to the functions $M_{x}\left(\phi_{\mathrm{p}},\beta\right)$,
$M_{y}\left(\phi_{\mathrm{p}},\beta\right)$ and $M_{xy}\left(\phi_{\mathrm{p}},\beta\right)$
characterised by quadratic trigonometric factors and \emph{does not}
permit the implementation of all logic gates. Achieving universality
when starting from $\rho_{x}$ therefore requires a different rotation
operator that never commutes with $\rho_{x}$. This could be achieved
by choosing different pairs of parameters from the set of parameters
describing the single-pulse experiment (Figure~\ref{fig:NMR-pulse-experiment})
such that a non-commuting rotation operator for $\rho_{x}$ results.
Alternatively, one can keep the pair of parameters $\phi_{\mathrm{p}}$
and $\beta$, and use more than one r.f.\ pulse, with rotation operators
such as $R_{\phi_{\mathrm{p}2}}\left(\beta_{2}\right)R_{\phi_{\mathrm{p1}}}\left(\beta_{1}\right)$.
This latter option we consider next.

\subsubsection{Two-pulse gates}

The initial density matrix $\rho_{x}$ transforms under the influence
of two r.f.\ pulses $R_{\phi_{\mathrm{p}2}}\left(\beta_{2}\right)R_{\phi_{\mathrm{p1}}}\left(\beta_{1}\right)$
as
\begin{eqnarray}
\rho^{\left(x\right)}\left(\phi_{\mathrm{p2}},\beta_{2},\phi_{\mathrm{p1}},\beta_{1}\right)  =  R_{\phi_{\mathrm{p}2}}\left(\beta_{2}\right)R_{\phi_{\mathrm{p1}}}\left(\beta_{1}\right)\rho_{x}R_{\phi_{\mathrm{p1}}}^{\dagger}\left(\beta_{1}\right)R_{\phi_{\mathrm{p}2}}^{\dagger}\left(\beta_{2}\right)\label{eq:two-pulse-transformation}
\end{eqnarray}
Thus, the NMR detectable quantities $M_{x}\left(\phi_{\mathrm{p2}},\beta_{2},\phi_{\mathrm{p1}},\beta_{1}\right)$, \\
$M_{y}\left(\phi_{\mathrm{p2}},\beta_{2},\phi_{\mathrm{p1}},\beta_{1}\right)$
 and $M_{xy}\left(\phi_{\mathrm{p2}},\beta_{2},\phi_{\mathrm{p1}},\beta_{1}\right)$
are functions of the four variables $\phi_{\mathrm{p2}}$, $\beta_{2}$,
$\phi_{\mathrm{p1}}$ and $\beta_{1}$. Any two of these variables
can serve as logic gate input parameters with the remaining two
fixed. This offers a high degree of flexibility in assigning parameters
as variable (and to control the logic gate), and as fixed.

A trivial example is based on the fact that all 2-input gates can
be implemented when starting from thermal equilibrium magnetisation
by a single pulse. Starting from $\rho_{x}$, a first r.f.\ pulse
$R_{\frac{\pi}{2}}\left(-\frac{\pi}{2}\right)$ generates $\rho_{z}^{\mathrm{eq}}$
and yields $\phi_{\mathrm{p1}}=\frac{\pi}{2}$ and $\beta_{1}=-\frac{\pi}{2}$.
From there on any second r.f.\ pulse $R_{\phi_{\mathrm{p2}}}\left(\beta_{2}\right)$
generates a functional behaviour identical to that of the single-pulse
scenario in Eq.~\eqref{eq:thermal-eq-single-pulse}, generating universality.

\begin{figure}
\begin{centering}
\includegraphics[width=1\columnwidth]{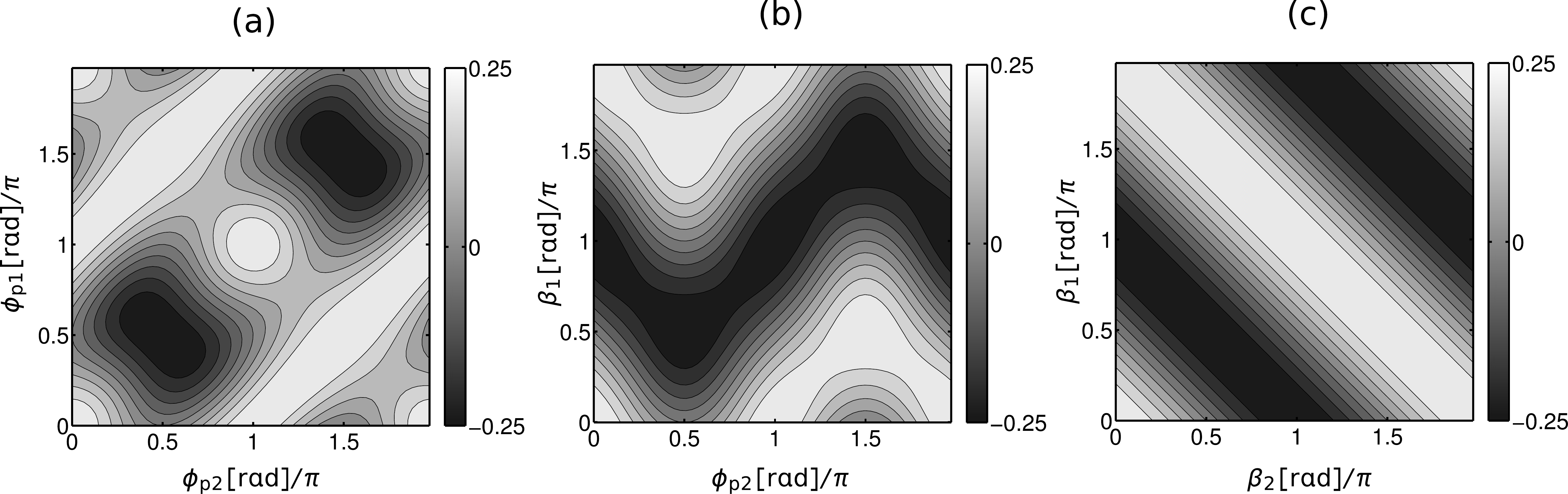}
\par\end{centering}

\caption{\label{fig:Contour-plots-two-pulse-Mx}Contour plots of $M_{x}\left(\phi_{\mathrm{p2}},\beta_{2},\phi_{\mathrm{p1}},\beta_{1}\right)$
corresponding to examples (a)--(c) in Table~\ref{tab:Example-two-pulse-Mx}.}
\end{figure}

Another strategy uses the commutation properties of the rotation operators.
In the single-pulse scenario, starting from $\rho_{z}^{\mathrm{eq}}$
all logic gates can be implemented because the overall rotation operator
never commutes with the initial density matrix. In the two-pulse scenario,
$R_{\phi_{\mathrm{p}2}}\left(\beta_{2}\right)R_{\phi_{\mathrm{p1}}}\left(\beta_{1}\right)$
must not commute with $\rho_{x}$. This implies further that the first
r.f.\ pulse $R_{\phi_{\mathrm{p1}}}\left(\beta_{1}\right)$ must
not commute with $\rho_{x}$, otherwise Eq.~\eqref{eq:two-pulse-transformation}
simplifies to $\rho^{\left(x\right)}\left(\phi_{\mathrm{p2}},\beta_{2}\right)=R_{\phi_{\mathrm{p}2}}\left(\beta_{2}\right)\rho_{x}R_{\phi_{\mathrm{p}2}}^{\dagger}\left(\beta_{2}\right)$
(identical to Eq.~\eqref{eq:single-pulse-tranformation-superposition})
which \emph{can not} implement all logic gates. To avoid commutation,
$\phi_{\mathrm{p1}}$ should never be fixed to values $n\pi$
(Eq.~\eqref{eq:singel-pulse-commutator-superposition}). The second
r.f.\ pulse $R_{\phi_{\mathrm{p}2}}\left(\beta_{2}\right)$ must
not commute with the spin basis operators $I_{x}$ or $I_{y}$ when
calculating $M_{x}\left(\phi_{\mathrm{p2}},\beta_{2},\phi_{\mathrm{p1}},\beta_{1}\right)=\overline{\Braket{I_{x}}}$
or $M_{y}\left(\phi_{\mathrm{p2}},\beta_{2},\phi_{\mathrm{p1}},\beta_{1}\right)=\overline{\Braket{I_{y}}}$ 
 (Eq.~\eqref{eq:Overall-Magnetisation}), otherwise\\ $R_{\phi_{\mathrm{p}2}}\left(\beta_{2}\right)R_{\phi_{\mathrm{p1}}}\left(\beta_{1}\right)\rho_{x}R_{\phi_{\mathrm{p1}}}^{\dagger}\left(\beta_{1}\right)I_{x}R_{\phi_{\mathrm{p}2}}^{\dagger}\left(\beta_{2}\right)$
is just a similarity transform of $R_{\phi_{\mathrm{p1}}}\left(\beta_{1}\right)\rho_{x}R_{\phi_{\mathrm{p1}}}^{\dagger}\left(\beta_{1}\right)I_{x}$
under which the trace is invariant. A commutation $\left[R_{\phi_{\mathrm{p}2}}\left(\beta_{2}\right),I_{x}\right]=0$
can be avoided by setting $\phi_{\mathrm{p}2}\neq n\pi$.
For $M_{y}\left(\phi_{\mathrm{p2}},\beta_{2},\phi_{\mathrm{p1}},\beta_{1}\right)$
a commutator $\left[R_{\phi_{\mathrm{p}2}}\left(\beta_{2}\right),I_{y}\right]=0$
is avoided for $\phi_{\mathrm{p}}\neq \left(2n+1\right)\frac{\pi}{2}$
. These constraints on the parameters of type phase $\phi_{\mathrm{p}}$
implement the functions $M_{x}\left(\phi_{\mathrm{p2}},\beta_{2},\phi_{\mathrm{p1}},\beta_{1}\right)$
such as those shown in Table~\ref{tab:Example-two-pulse-Mx} and
Figure~\ref{fig:Contour-plots-two-pulse-Mx}: these examples were
deliberately chosen such that the two fixed variables are taken as
having equal values of $\frac{\pi}{2}$.

\begin{table}
	\centering
	\footnotesize
\begin{tabular}{cccccl}
\toprule
 & $\phi_{\mathrm{p2}}$ & $\beta_{2}$ & $\phi_{\mathrm{p1}}$ & $\beta_{1}$ & $M_{x}\left(\phi_{\mathrm{p2}},\beta_{2},\phi_{\mathrm{p1}},\beta_{1}\right)$\tabularnewline
\midrule
(a) & $A$ & $\frac{\pi}{2}$ & $B$ & $\frac{\pi}{2}$ & $\frac{\lambda_{\mathrm{B}}}{4}\left(\cos\phi_{\mathrm{p2}}\cos\phi_{\mathrm{p1}}\cos\left(\phi_{\mathrm{p2}}-\phi_{\mathrm{p1}}\right)-\sin\phi_{\mathrm{p2}}\sin\phi_{\mathrm{p1}}\right)$\tabularnewline
(b) & $A$ & $\frac{\pi}{2}$ & $\frac{\pi}{2}$ & $B$ & $\frac{\lambda_{\mathrm{B}}}{4}\left(\cos^{2}\phi_{\mathrm{p2}}\cos\beta_{1}-\sin\phi_{\mathrm{p2}}\sin\beta_{1}\right)$\tabularnewline
(c) & $\frac{\pi}{2}$ & $A$ & $\frac{\pi}{2}$ & $B$ & $\frac{\lambda_{\mathrm{B}}}{4}\cos\left(\beta_{2}+\beta_{1}\right)$\tabularnewline
(d) & $\frac{\pi}{2}$ & $A$ & $B$ & $\frac{\pi}{2}$ & $\frac{\lambda_{\mathrm{B}}}{4}\left(\cos^{2}\phi_{\mathrm{p1}}\cos\beta_{2}-\sin\phi_{\mathrm{p1}}\sin\beta_{2}\right)$\tabularnewline
(e) & $A$ & $B$ & $\frac{\pi}{2}$ & $\frac{\pi}{2}$ & $-\frac{\lambda_{\mathrm{B}}}{4}\sin\phi_{\mathrm{p2}}\sin\beta_{2}$\tabularnewline
(f) & $\frac{\pi}{2}$ & $\frac{\pi}{2}$ & $A$ & $B$ & $-\frac{\lambda_{\mathrm{B}}}{4}\sin\phi_{\mathrm{p1}}\sin\beta_{1}$\tabularnewline
\bottomrule
\end{tabular}
\caption{\label{tab:Example-two-pulse-Mx}Examples of logic gate implementations
from $M_{x}\left(\phi_{\mathrm{p2}},\beta_{2},\phi_{\mathrm{p1}},\beta_{1}\right)$
using any possible pair of variables as gate input $A$ and $B$,
and fixing the remaining two parameters to $\frac{\pi}{2}$. Apart
from a change in sign, examples (e) and (f) are identical to the case
in Figure~\ref{fig:Single-pulse-measurments-Iz}(a). Example (d)
is identical to (b).}
\end{table}
 
 One can see that class 0, 1 and 3 gates can be implemented by all
possible permutations of fixed pairs of variables of identical value
(here $\frac{\pi}{2}$). The non-commutation constraint provides the
non-canalising behaviour of class 3 gates, while the inherent periodicity
of the system is sufficient to provide the canalising input configurations
necessary for class 0 and 1 gates. However, periodicity does not suffice
for a class 2 gate implementation. Comparing the single-pulse scenario
(Figure~\ref{fig:Single-pulse-measurments-Iz}(a)) with the contour
plots in Figure~\ref{fig:Contour-plots-two-pulse-Mx} demonstrates
that class 2 gate implementations require horizontal and vertical
traces of constant value zero. Such traces are absent in the contour
plots in Figure~\ref{fig:Contour-plots-two-pulse-Mx}. In short, arbitrarily
choosing pairs of fixed parameters as equal valued (not just for $\frac{\pi}{2}$)
leads to the loss of universality.

\begin{figure}
\begin{centering}
\includegraphics[width=0.8\columnwidth]{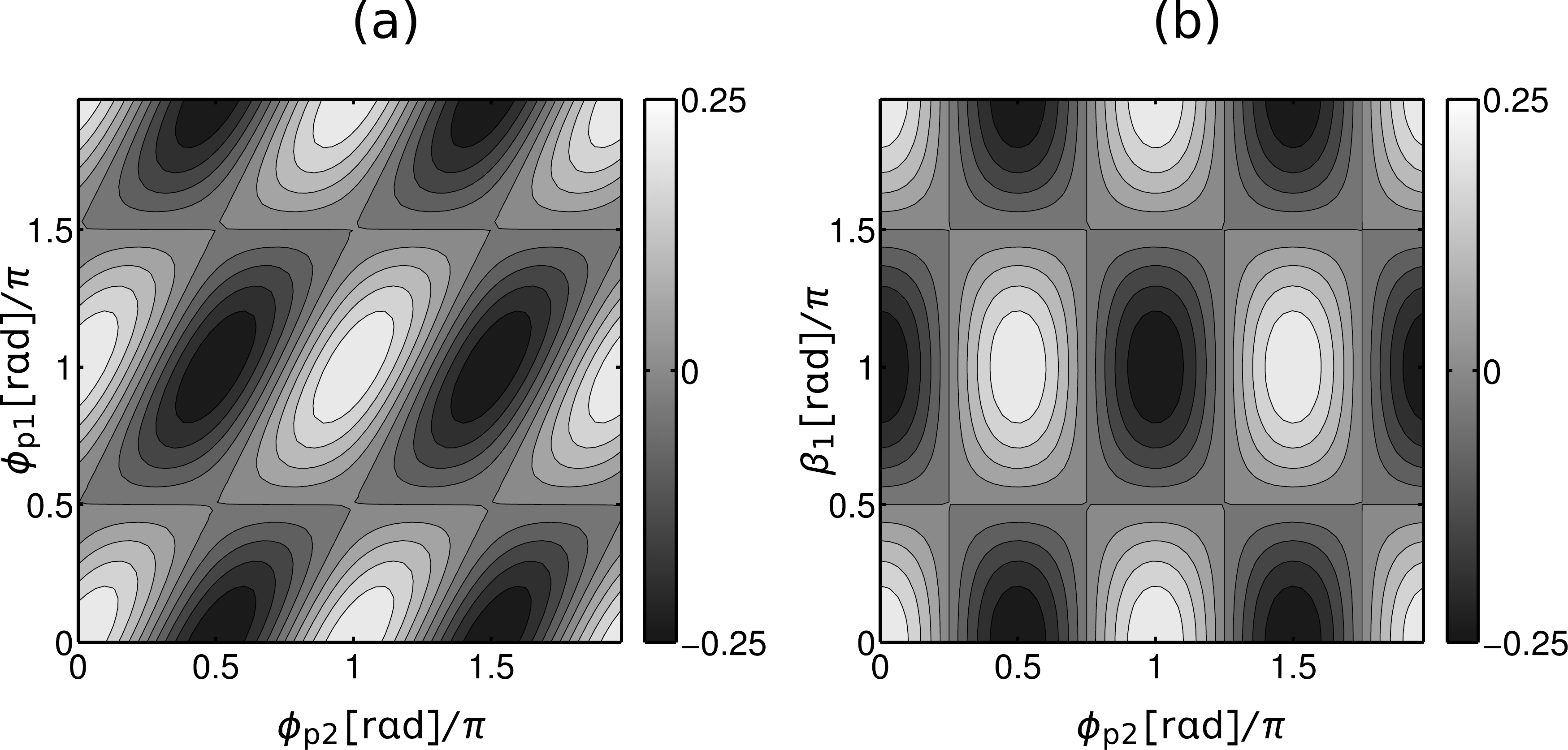}
\par\end{centering}

\caption{\label{fig:Contour-plot-of-better-beta}Contour plot of $M_{x}\left(\phi_{\mathrm{p2}},\beta_{2},\phi_{\mathrm{p1}},\beta_{1}\right)$
for (a) $\beta_{1}=\frac{\pi}{2}$, $\beta_{2}=\pi$ and for (b) $\phi_{\mathrm{p1}}=\frac{\pi}{2}$,
$\beta_{2}=\pi$.}

\end{figure}

Let us examine if universality is regained if the fixed parameters
are no longer taken as equal valued. For example, taking $\beta_{1}=\frac{\pi}{2}$
and $\beta_{2}=\pi$ gives $M_{x}\left(\phi_{\mathrm{p2}},\beta_{2},\phi_{\mathrm{p1}},\beta_{1}\right)=\frac{\lambda_{\mathrm{B}}}{4}\cos\left(2\phi_{\mathrm{p2}}-\phi_{\mathrm{p1}}\right)\cos\left(\phi_{\mathrm{p1}}\right)$.
The corresponding contour plot is shown in Figure~\ref{fig:Contour-plot-of-better-beta}(a)
highlighting the presence of constant horizontal zero-valued traces.
Allowing for unequal-valued pairs of parameters $\beta_{1}$ and $\beta_{2}$
(or $\phi_{\mathrm{p1}}$ and $\phi_{\mathrm{p2}}$ ) permits implementation
of class 0, 1 and 2 gates but not class 3. 

Universality is regained if pairs of parameters of \emph{different}
\emph{types}, e.g. $\phi_{\mathrm{p1}}$ and $\beta_{2}$ are fixed
and do \emph{not} assume equal values. Figure~\ref{fig:Contour-plot-of-better-beta}(b)
shows that $\phi_{\mathrm{p1}}=\frac{\pi}{2}$, $\beta_{2}=\pi$ gives
$M_{x}\left(\phi_{\mathrm{p2}},\beta_{2},\phi_{\mathrm{p1}},\beta_{1}\right)=\frac{\lambda_{\mathrm{B}}}{4}\cos2\phi_{\mathrm{p2}}\cos\beta_{1}$.
Both horizontal and vertical zero-valued traces are re-established
in the corresponding contour plot (Figure~\ref{fig:Contour-plot-of-better-beta}(b));
all 2-input logic gates can be obtained.

The flexibility and the ease of implementation as seen from our illustrative
examples are good indicators for the richness of the natural computational
potential of this system. Computational operations more complicated
than just 2-input boolean logics are therefore well within the grasp
of this system. This includes multi-input logic gates, continuous
logic, and analogue computing implementations. The ubiquitous occurrence
of function values $\left\{ -1,0,1\right\} $, due to the trigonometric
system functions, especially holds promise for the implementation
of a balanced ternary logic as being natural to this particular system
\cite{Bechmann2011}. Note also, that we can reverse the operation
of a logic gate if we know the pulse(s) that originally generated
the output \cite{Levitt2008}. This does not match with the usual
predictions about a computation where e.g.\ NAND gates are not reversible
\cite{Bennett1982}. Obviously our raw output stores additional information 
that can be exploited for more sophisticated computations.

\section{Summary and Conclusions}

We have provided a design approach to analysing novel substrates in
order to determine which of their parameters can be used to implement
boolean logic gates. We have illustrated this with a case study drawn
from NMR-based classical computation. The design process requires
cataloguing the parameters that are naturally used to describe and
manipulate the target system, analysing their behaviours in combinations,
and then matching the resulting patterns of behaviour with the corresponding
behaviour patterns of the target gate classes. Our case study here
focusses on two particular parameters ($\beta=\kappa_{\mathrm{p}}\tau_{\mathrm{p}}$
and $\phi_{\mathrm{p}}$); a full design study would assess other
combinations of other parameters, since the aim is not simply to find
\textit{some} solution, but to analyse the `natural' computational
capabilities of the substrate. For example, in our work on continuous
gates \cite{BechmannICES2010}, we focussed on $\omega_{\mathrm{p}}$
and $\tau_{\mathrm{p}}$.

The design approach is not restricted to 2-input gates; combining
several parameters can produce patterns corresponding to more complex
gates. More sophisticated experiments could directly exploit symmetry 
properties of the Hamiltonian (Eq.~\eqref{eq:Htimeindep}--\eqref{eq:Ucyclic}).
Eventually, however, specific behaviours will be more easily
achieved by combining simpler gates in circuits than by directly designing
complex gates.

Circuit design requires additional analysis to determine how individual
gates can be combined in a circuit in a manner `natural' for the substrate
in question. Circuit design requires determining techniques for: sequencing
parameter manipulations to implement the sequence of gate operations
in a circuit; combining manipulations to implement multiple gate operations
in parallel; routing and transforming the output of one gate to the
appropriate input of the next. We will address these circuit design
issues in a companion paper \cite{Bechmann-wiring}.

This work addresses only how to implement `classical' boolean logic
gates in unconventional substrates, yet it is clear that the example
system has further information available, that is thrown away when
viewing it as a boolean gate. If this extra information is instead
retained and exploited, more powerful computation becomes available
\cite{Abbott2011}. The real computational power of these novel substrates
will come from not viewing them as merely alternative ways of implementing
classical logic gates, but from exploiting them to implement non-classical
forms of computation: quantum, continuous (analogue), hybrid, and
more. The design approach described here forms a first step in a principled
approach for analysing substrates with a view to performing a specified
form of computation.

\section{Acknowledgements}

Support of this work by the Deutsche Forschungsgemeinschaft and Leverhulme
grant FM/00224/AM is gratefully acknowledged. We thank Alastair Abbott
for constructive comments.

\bibliography{StrongWeak}

\begin{thebibliography}{10}

\bibitem{Abbott2011}
Alastair~A. Abbott, Matthias Bechmann, Christian~S. Calude, and Angelika
  Sebald.
\newblock (May 2011).
\newblock A nuclear magnetic resonance implementation of a classical
  {Deutsch-Jozsa} algorithm.
\newblock submitted.

\bibitem{Adamatzky_2007}
Andrew Adamatzky.
\newblock (2007).
\newblock Physarum machines: encapsulating reaction-diffusion to compute
  spanning tree.
\newblock {\em Naturwissenschaften}, 94(12):975--980.

\bibitem{Adleman11111994}
Leonard~M. Adleman.
\newblock (1994).
\newblock Molecular computation of solutions to combinatorial problems.
\newblock {\em Science}, 266(5187):1021--1024.

\bibitem{Amos}
Martyn Amos.
\newblock (2005).
\newblock {\em Theoretical and Experimental DNA Computation}.
\newblock Springer.

\bibitem{Bechmann-wiring}
Matthias Bechmann, Angelika Sebald, and Susan Stepney.
\newblock Direct wiring of multi-gate {NMR} logic circuits.
\newblock In preparation.

\bibitem{BechmannICES2010}
Matthias Bechmann, Angelika Sebald, and Susan Stepney.
\newblock (2010).
\newblock From binary to continuous gates---and back again.
\newblock In {\em ICES 2010, York, UK, September 2010}, volume 6274 of {\em
  LNCS}, pages 335--347. Springer.

\bibitem{Bechmann2011}
Matthias Bechmann, Angelika Sebald, and Susan Stepney.
\newblock (2011).
\newblock Design of ternary logic gates: a nucler magnetic resonance case
  study.
\newblock In preparation.

\bibitem{Bennett1982}
Charles~H. Bennett.
\newblock (1982).
\newblock The thermodynamics of computation---a review.
\newblock {\em Int. J. Theor. Phys.}, 21:905--940.

\bibitem{Correia2001}
Vin\'icius~P. Correia, Andr\'e~I. Reis, Cep Porto, and Alegre~Rs Brasil.
\newblock (2001).
\newblock Classifying n-input boolean functions.
\newblock In {\em Proc. VII Workshop de Iberchip, IWS, Montevideo}, page~58.

\bibitem{Drossel2009}
Barbara Drossel.
\newblock (2009).
\newblock {\em Random Boolean Networks}, pages 69--110.
\newblock Wiley.

\bibitem{Golomb1959}
Solomon Golomb.
\newblock (1959).
\newblock On the classification of boolean functions.
\newblock {\em IRE Transactions on Information Theory}, 5(5):176--186.

\bibitem{Harding}
Simon~L. Harding and Julian~F. Miller.
\newblock (2004).
\newblock A tone discriminator in liquid crystal.
\newblock In {\em CEC 2004}, pages 1800--1807. IEEE Press.

\bibitem{Miller}
Simon~L. Harding and Julian~F. Miller.
\newblock (2005).
\newblock Evolution {\sl in materio}: A real-time robot controller in liquid
  crystal.
\newblock In {\em Proc. NASA/DoD Conference on Evolvable Hardware}, pages
  229--238. IEEE Press.

\bibitem{Harding06}
Simon~L. Harding, Julian~F. Miller, and Edward~A Rietman.
\newblock (2008).
\newblock Evolution in materio: Exploiting the physics of materials for
  computation.
\newblock {\em Int. J. Unconventional Computing}, 4(2):155--194.

\bibitem{Hardy2001}
Yorik Hardy and Willi~H. Steeb.
\newblock (2001).
\newblock {\em Classical and Quantum Computing}.
\newblock Birkh\"auser Verlag, Basel.

\bibitem{Jones2001}
Jonathan~A. Jones.
\newblock (2001).
\newblock {NMR} quantum computation.
\newblock {\em Prog. Nucl. Magn. Reson. Spectrosc.}, 38(4):325--360.

\bibitem{Jones2011}
Jonathan~A. Jones.
\newblock (2011).
\newblock Quantum computing with nmr.
\newblock {\em Progress in Nuclear Magnetic Resonance Spectroscopy},
  59(2):91--120.

\bibitem{SS-UC11}
Viv Kendon, Angelika Sebald, Susan Stepney, Matthias Bechmann, Peter Hines, and
  Robert~C. Wagner.
\newblock (2011).
\newblock Heterotic computing.
\newblock In {\em Unconventional Computation 2011}, number 6714 in LNCS, pages
  113--124. Springer.

\bibitem{Kuhnert_1989}
Lothar Kuhnert, Konstantin Agladze, and Valentin Krinsky.
\newblock (1989).
\newblock Image processing using light-sensitive chemical waves.
\newblock {\em Nature}, 337:244--247.

\bibitem{Levitt2008}
Malcolm.~H. Levitt.
\newblock (April 2008).
\newblock {\em Spin Dynamics: Basics of nuclear magnetic resonance}.
\newblock John Wiley \& Sons, Ltd, Chichester, 2nd edition.

\bibitem{Mills_2008}
Jonathan~W. Mills.
\newblock (2008).
\newblock The nature of the extended analog computer.
\newblock {\em Phys. D}, 237(9):1235--1256.

\bibitem{Mills_al_2006}
Jonathan~W. Mills, Matt Parker, Bryce Himebaugh, Craig Shue, Brian Kopecky, and
  Chris Weilemann.
\newblock (2006).
\newblock ``{Empty Space}'' computes: The evolution of an unconventional
  supercomputer.
\newblock In {\em Proc. of the 3rd conference on Computing frontiers}, pages
  115--126.

\bibitem{Motoike2005107}
Ikuko~N. Motoike and Andrew Adamatzky.
\newblock (2005).
\newblock Three-valued logic gates in reaction-diffusion excitable media.
\newblock {\em Chaos, Solitons \& Fractals}, 24(1):107--114.

\bibitem{Rosello-Merino2010}
Marta Rosell{\'o}-Merino, Matthias Bechmann, Angelika Sebald, and Susan
  Stepney.
\newblock (2010).
\newblock Classical computing in nuclear magnetic resonance.
\newblock {\em Int. J. Unconventional Computing}, 6(3--4):163--195.

\bibitem{Sakurai1994}
Jun~J. Sakurai.
\newblock (1994).
\newblock {\em Modern Quantum Mechanics}.
\newblock Addison-Wesley Publishing Company, Reading, MA, {R}evised edition.

\bibitem{Schumacher1995}
Benjamin Schumacher.
\newblock (Apr 1995).
\newblock Quantum coding.
\newblock {\em Phys. Rev. A}, 51(4):2738--2747.

\bibitem{Sielewiesiuk_2001}
Jakub Sielewiesiuk and Jerzy Gorecki.
\newblock (2001).
\newblock Logical functions of a cross-junction of excitable chemical media.
\newblock {\em J. Phys. Chem. A}, 105(35):8189--8195.

\bibitem{Slepian1953}
David Slepian.
\newblock (1953).
\newblock On the number of symmetry types of boolean functions of $n$
  variables.
\newblock {\em Can. J. Math.}, 5(2):185--193.

\bibitem{SS-PhysicaD-08}
Susan Stepney.
\newblock (July 2008).
\newblock The neglected pillar of material computation.
\newblock {\em Physica D}, 237(9):1157--1164.

\bibitem{SS-BCS-VCS}
Susan Stepney, Samson Abramsky, Andy Adamatzky, Colin Johnson, and Jon Timmis.
\newblock (2008).
\newblock Grand challenge 7: Journeys in non-classical computation.
\newblock In {\em Visions of Computer Science}, pages 407--421. BCS.

\bibitem{Tero22012010}
Atsushi Tero, Seiji Takagi, Tetsu Saigusa, Kentaro Ito, Dan~P. Bebber, Mark~D.
  Fricker, Kenji Yumiki, Ryo Kobayashi, and Toshiyuki Nakagaki.
\newblock (2010).
\newblock Rules for biologically inspired adaptive network design.
\newblock {\em Science}, 327(5964):43--9--442.

\bibitem{Toth_1995}
{\'A}gota T{\'o}th and Kenneth Showalter.
\newblock (1995).
\newblock Logic gates in excitable media.
\newblock {\em J. Chem. Phys.}, 103:2058--2066.

\bibitem{Woods2008}
Damien Woods and Thomas~J. Naughton.
\newblock (2008).
\newblock Parallel and sequential optical computing.
\newblock In {\em Optical SuperComputing}, volume 5172 of {\em LNCS}, pages
  70--86. Springer.

\bibitem{Woods2009}
Damien Woods and Thomas~J. Naughton.
\newblock (2009).
\newblock Optical computing.
\newblock {\em Appl. Math. Comput.}, 215(4):1417--1430.

\bibitem{Yau1970}
Stephen~S. Yau and Calvin~K. Tang.
\newblock (feb. 1970).
\newblock Universal logic modules and their applications.
\newblock {\em IEEE Transactions on Computers}, C-19(2):141--149.

\end{thebibliography}
\end{document}